\documentclass[twocolumn,showpacs,preprintnumbers,amsmath,amssymb]{revtex4}
\usepackage{graphicx}% Include figure files
\usepackage{dcolumn}% Align table columns on decimal point
\usepackage{bm}% bold math

\begin{document}

%\preprint{APS/*!*!*}

\title{Spin magnetization of strongly correlated
electron gas confined in a two-dimensional finite lattice}

\author{M. Ni\c{t}\u{a}, V. Dinu, and A. Aldea}
%\affiliation{Institute of Physics and Technology of Materials, POBox 
\address{Institute of Physics and Technology of Materials, POBox 
MG7, Bucharest-Magurele, Romania}
\author{B. Tanatar}
%\affiliation{Department of Physics, Bilkent University, 06533 Bilkent, 
\address{Department of Physics, Bilkent University, 06533 Bilkent, 
Ankara, Turkey}
%\date{\today}
%\maketitle
\begin{abstract}
The influence of disorder and interaction on
the ground state polarization of the two-dimensional (2D) correlated 
electron gas is studied by numerical investigations of unrestricted 
Hartree-Fock equations. The ferromagnetic ground state is found to be 
plausible when the electron number is lowered and the interaction and 
disorder parameters are suitably chosen. For a finite system at 
constant electronic density the disorder induced spin polarization is 
cut off when the electron orbitals become strongly localized to the 
individual network sites. The fluctuations of the interaction matrix 
elements are calculated and brought out as favoring the ferromagnetic 
instability in the extended and weak localization regime.
The localization effect of the Hubbard interaction term is discussed.

\end{abstract}

\maketitle

\section{Introduction}

The combined effect of carrier interactions and Pauli principle leads 
to itinerant ferromagnetism in metals whenever the kinetic energy gain 
by parallel alignment of electronic spins is exceeded by the
exchange energy of the antisymmetric wave function (\cite{kim}). 
The simplest model is the Hund's rule in atoms, where the energy is 
minimized when the fermionic wave function corresponds to the alignment 
of spins.
%In general, whenever the exchange energy between two close states is
%equal to level separation, a spin polarization transition is 
%possible (Stoner instability) (\cite{kim}).
Meanwhile the electron-spin correlations seems to be an important 
factor in the two-dimensional (2D) metal-insulator 
transition \cite{mit}. Measurements of the in-plane magnetoconductivity
of a dilute 2D electron system in silicon heterostructures have
shown the evidence for a zero-temperature quantum phase transition at
critical density, indicating the existence of a ferromagnetic 
instability (\cite{vitkalov},\cite{shashkin}).
On the other hand a new method that measures the minute thermodynamic
spin magnetization of a dilute 2D electron system in a silicon inversion
layer favors the paramagnetic phase over the spontaneous magnetization
\cite{prus}.

The interaction strength between electrons is traditionally described 
by the 2D Wigner-Seitz radius,
$r_{s}=1/(\pi n_{s}a_B^2)^{1/2}$, where $a_{B}$ is the Bohr radius
and $n_{s}=n/L^{2}$ sheet density of electrons.
% In the
%single electron valley, it is equal to ratio between Coulomb energy to
%Fermi energy $r_{s}=U/(t\sqrt{\pi n_{s}})$.
As a function of $r_{s}$ the ground state of the 2D electron
gas can be, for instance,
a Wigner crystal (in dilute limit, at $r_{s} \ge 35$ \cite{tanatar}),
a ferromagnetic Fermi liquid with spontaneous magnetization or a 
paramagnetic Fermi liquid. 
An improved hypernetted-chain approximation was employed to study the 
2D electron liquid at full Fermi degeneracy \cite{perrot,bulutay}. 
The results indicate the 
paramagnetic phase as the ground state of the system until the Wigner 
crystallization density, even the separation with polarized 
states become minute (the energy difference with the ferromagnetic phase
diminishes to  milliRydbergs). In contrast, the transition from 
paramagnetic fluid phase to Wigner 
crystallization was recently studied for a 2D electron gas by 
diffusion Monte Carlo simulations including
the backflow corrections \cite{varsano} showing
the stability range of a polarized phase in-between. The earlier work 
including the backflow correlations did not find
the polarization transition \cite{kwon}. More recently a weakly 
first-order transition to a spin-polarized state was found to occur 
shortly before the Wigner crystallization \cite{attaccalite}.

Hubbard model is the simplest tool that can capture some of the salient
properties of correlated systems. A delocalized Coulomb phase in
2D was observed in the calculations of by Waintal 
{\it et al}.\cite{waintal} which showed the 
transition from the Anderson insulator towards an extended phase. 
Mean-field approximation considered in an unrestricted Hartree-Fock 
ansatz (UHF) was compared with constrained path quantum Monte Carlo 
method for repulsive Hubbard model at 
half-filling\cite{enjarlan} and found both Anderson and Mott insulator
phases.
In a 3D system numerical mean-field studies by Tusch and
Logan\cite{tusch} at the unrestricted 
Hartree-Fock level revealed the phase diagram at half-filling. 
The phase diagram in a 2D system is studied by Hirsch\cite{hirsch} using
a square lattice with nearest-neighbor hopping.

Nowadays there is also a growing interest in the spin 
polarization and Stoner instability of finite-size disordered systems 
\cite{baranger},\cite{jacquod},\cite{eisenberg},\cite{brouwer},
\cite{andreev}.
%(\cite{baranger},\cite{jacquod},\cite{eisenberg},\cite{brouwer},\cite{andreev})
In quantum dots, the singlet-triplet
transitions even for weak interactions cause switching between states
and a kink in the conductance \cite{baranger}. For interaction 
strength below the Stoner criterion a nonzero spin ground state was
found for an ensemble of small metallic grains in the
Hartree-Fock approximation \cite{brouwer}. Close to the Stoner 
instability in a 2D metallic system there is an exponentially small 
probability for the appearance of local spin droplets\cite{narozhny}.
The tendency of disorder to exhibit the magnetism of ground state 
was already pointed out in some recent works\cite{eisenberg},
\cite{benenti},\cite{berkovits}.
Meanwhile the fluctuations of the interaction matrix for different 
disorder ranges seem to have an important role in the spin polarization 
transition for finite systems. At crossover between random matrix 
ensembles the correlations between different eigenvectors give
rise to enhanced fluctuations of the interaction matrix elements (IME) 
in small metallic grains and semiconductor quantum dots \cite{adam}.
When their fluctuations dominate a 2D random interaction model
predicts an increased  probability of a minimal spin for the ground 
state lowering in the tails of the energy band \cite{jacquod} and for 
a two-orbital model they also help in stabilizing a ground state with 
minimal spin \cite{hirose}.

Our goal in this work is to study the influence of disorder and 
interactions on the polarization of the ground state for a 2D 
finite lattice. We treat the interaction effects at the unrestricted 
Hartree-Fock level (UHF), neglecting the off-diagonal contribution of 
the Coulomb matrix. This approximation permits us to write an effective
one-body Hamiltonian, suitable for calculating the conductances in the 
Landauer-B\"{u}ttiker formalism \cite{buttiker}. We solve numerically 
the self-consistent set of equations for 2D systems with Hubbard 
interaction grater than the Coulomb interaction strength ($U_{H}\gg U$).
For different disorder ranges we study the fluctuations of the 
interaction matrix elements and point out their role 
in the disorder induced spin polarization.

The rest of this paper is organized as follows. In the next
section we outline the theoretical framework of unrestricted
Hartree-Fock approximation and develop an effective Hamiltonian
for the problem at hand. In Sec.\,III we present our results
for the fluctuations in the interaction matrix elements in
various disorder regimes. Section IV discusses the possibility
of disorder enhanced spin polarization within our model.
We close by listing our main conclusions in Sec.\,V.

\section{The unrestricted Hartree-Fock formalism}

We study a disordered rectangular lattice with $n$ fermions on
 $L^{2}=L_{x}\cdot L_{y}$ sites
with vanishing boundary conditions imposed.
If $c_{i\sigma}^{\dagger}$ and $c_{i\sigma}$ are the creation and
annihilation operators of an electron at site $i$ with spin $\sigma$, 
the Hamiltonian will be defined as follows:
% \cite{mahan}):
\begin{eqnarray}\label{}
H=&&t\sum_{<i,j>,\sigma}c_{i\sigma}^{\dagger}c_{j\sigma}
+\sum_{i,\sigma}\omega_{i}n_{i,\sigma}
+U_{Hub}\sum_{i}n_{i\uparrow}n_{i\downarrow}
\nonumber\\
&&+U\sum_{i\ne j,\sigma,\sigma'}\frac{n_{i,\sigma}n_{j,\sigma'}}{|i-j|}
\end{eqnarray}
where $n_{i,\sigma} = c_{i\sigma}^{\dagger}c_{i\sigma}$ is the 
occupation number operator, $\omega_{i}$ is the random Anderson 
disorder ($\omega_{i}\in[-W/2,W/2]$), $U_{Hub}$ and $U$ are the 
strength of the Hubbard and long-range Coulomb interaction, and $t$ 
is the energy unit of the hopping integral (it will be considered 
equal to unity). 
For $\omega_{i}$ set to zero, the Hamiltonian of Eq.\,(1) is the 
extended Hubbard model, used for studying the correlated 
systems\cite{mahan}. 
The Hubbard term was first introduced in \cite{hubbard} and the 
argument in favor of it is that only electrons with opposite spins can 
occupy the same state (Pauli exclusion principle).

We write the Hamiltonian given in Eq.\,(1) in a basis formed by 
two sets of orthonormalized single-particle states 
$\phi_{\alpha}^{\uparrow}(i)$, $\phi_{\alpha}^{\downarrow}(i)$ with
$\alpha \in [1,L^{2}]$ (unrestricted Hartree-Fock orbitals). Creation 
and annihilation operators in the new basis
$c_{\alpha\sigma}^{+}$ and $c_{\alpha\sigma}$ are given by the
transformation 
$c_{\alpha,\sigma}^{\dagger} = \sum_{i}\phi_{\alpha}^{\sigma}(i)
c_{i,\sigma}^{\dagger}$ and
$c_{\alpha,\sigma} = \sum_{i}\phi_{\alpha}^{\sigma\star}(i)
c_{i,\sigma}$.
Employing the variational principle, we look for the self-consistent 
set of eigenvectors $\phi_{\alpha}^{\sigma}(i)$ that minimize the 
ground state energy $E_{G}$ for a given number of electrons $n$ and a 
definite component of spin along an arbitrary $z$ axis, $S_{z}$.
The ground state will be a Slater determinant 
$|\Psi_{G}\rangle$
that correspond to the $n_{\sigma}$ electrons in the states 
$\phi_{\alpha}^{\sigma}(i)$ for $\alpha = 1,...,n_{\sigma}$ and
$\sigma = \uparrow , \downarrow$ ($n_{\uparrow}$+$n_{\downarrow}$=$n$):

\begin{equation}\label{}
|\Psi_{G}\rangle =
\prod_{{\alpha =1}}^{n_{\uparrow}}c_{\alpha,\uparrow}^{\dagger}
\prod_{{\alpha =1}}^{n_{\downarrow}}c_{\alpha,\downarrow}^{\dagger}
                |\Psi_{0}\rangle
\end{equation}
where $|\Psi_{0}\rangle$ is the vacuum \cite{gutzwiller}.
The eigenvectors $\phi_{\alpha}^{\sigma}(i)$
and the corresponding Lagrange multipliers $\epsilon_{\alpha}^{\sigma}$
that give the lowest energy solution for the ground state energy 
$E_{G}=\langle\Psi_{G}|H|\Psi_{G}\rangle$
will be calculated by the variational principle.
After a straightforward calculation one obtains
the following spin-separable Hamiltonian 
$H^{eff}=\sum_{\sigma}H_{\sigma}$ for $\sigma = \uparrow , \downarrow$:
%hamiltonian effective
\begin{eqnarray}
H_{\sigma}=&&\sum_{<i,j>}c_{i\sigma}^{\dagger}c_{j\sigma}
+\sum_{i}[ \omega_{i}+ V_{D}(i)+
V_{Hub}^{\sigma}(i)] n_{i,\sigma}\nonumber\\
&&-\sum_{i\ne j} V_{Ex}^{\sigma}(i,j) c_{i\sigma}^{\dagger}c_{j\sigma}
\end{eqnarray}
with  $\phi_{\alpha}^{\sigma}(i)$ and $\epsilon_{\alpha}^{\sigma}$
eigenvectors and eigenvalues of Eq.\,(3):
$H_{\sigma}|\phi_{\alpha}^{\sigma}\rangle=\epsilon_{\alpha}^{\sigma}|
\phi_{\alpha}^{\sigma}\rangle$.
The time reversal symmetry of $H$ ensures the possibility of choosing 
a real system of eigenvectors.

The matrix elements $V_{D}(i)$, $V_{Ex}^{\sigma}(i,j)$ 
and $V_{Hub}^{\sigma}(i)$ 
(direct, exchange and Hubbard interactions, respectively) are
related to the single particle eigenvectors by the following formulae 
(we stress that in the following $\overline{\sigma}$ means the opposite
spin of $\sigma$):
%elemente potential in hamiltonian effective
\begin{equation}\label{}
V_{D}(i)=2U\sum_{j\ne i} \frac{\sum_{\sigma}
\sum_{\alpha =1}^{n_{\sigma}} 
|\phi_{\alpha}^{\sigma}(j)|^{2}}{|i-j|}
\end{equation}

\begin{equation}\label{}
V_{Ex}^{\sigma}(i,j)=2U \frac{\sum_{\alpha =1}^{n_{\sigma}} 
\phi_{\alpha}^{\sigma*}(j) \phi_{\alpha}^{\sigma}(i)}   {|i-j|}
\end{equation}

\begin{equation}\label{}
V_{Hub}^{\sigma}(i)=U_{Hub} \sum_{\alpha =1}^{n_{\overline{\sigma}}} 
|\phi_{\alpha}^{\overline{\sigma}}(i)|^{2}
\end{equation}

%groun state energy
The ground state energy is given by:
\begin{eqnarray}
E_{G}=&&\sum_{\alpha \sigma}\epsilon_{\alpha}^{\sigma} +
U\sum_{\alpha \sigma,\beta \sigma'} V_{\alpha \sigma,\beta \sigma'}
                                      ^{\alpha \sigma,\beta \sigma'}
-U\sum_{\alpha \sigma,\beta \sigma} V_{\alpha \sigma,\beta \sigma}
^{\beta \sigma,\alpha \sigma}\nonumber\\
&&+U_{Hub} \sum_{\alpha \sigma,\beta \overline{\sigma}}
C_{\alpha \sigma,\beta \overline{\sigma}}
    ^{\alpha \sigma,\beta \overline{\sigma}}
\end{eqnarray}
where the second and the third terms are the contributions of direct 
and exchange interaction, respectively, to the ground state energy, 
and the last term is the Hubbard interaction. The summations in Eq.\,(7)
are over the occupied states $|\alpha\sigma\rangle$ from Eq.\,(2).
The Coulomb and Hubbard interaction matrix elements are given
as
%interaction matrix
\begin{equation}\label{}
V_{\alpha \sigma,\beta \sigma'}
^{\gamma \sigma,\delta \sigma'}=\sum_{i\ne j}\frac
{\phi_{\alpha}^{\sigma*}(i)\phi_{\gamma}^{\sigma}(i)
\phi_{\beta}^{\sigma'*}(j)\phi_{\delta}^{\sigma'}(j)}
{|i-j|}\, ,
\end{equation}
%hubard interaction matrix
\begin{equation}\label{}
C_{\alpha \sigma,\beta \sigma'}
^{\gamma \sigma,\delta \sigma'}=\sum_{i}
{\phi_{\alpha}^{\sigma*}(i)\phi_{\gamma}^{\sigma}(i)
\phi_{\beta}^{\sigma'*}(i)\phi_{\delta}^{\sigma'}(i)}\, .
\end{equation}
Omitting the off-diagonal matrix elements ($(\alpha,\beta)\not=
(\gamma,\delta)$) in Eq.\,(7) yields the Hartree-Fock approximation.
The reason for doing this will be discussed subsequently.

The basic parameters of our theoretical framework are the electron
number $n$, the long-range Coulomb interaction $U$, 
the Hubbard interaction strength $U_{Hub}$, and the disorder
amplitude $W$. Equations (3-7) constitute
a set of self-consistent equations to yield the 
eigenstates $\phi_{\alpha}^{\sigma},\epsilon_{\alpha}^{\sigma}$ and the
energy $E_{G}$ for a given set of spin occupation numbers 
($n_{\uparrow},n_{\downarrow}$) with $n_{\uparrow}+n_{\downarrow}=n$. 
At every step of the numerical calculation,
the matrix elements [described by Eqs.\,(4-6)] of $H^{eff}$ will be 
calculated for the spin occupation numbers that give the minimum 
energy in the previous step.
The first Slater determinant of Eq.\,(3) will be formed by 
noninteracting eigenfunctions of Hamiltonian [Eq.\,(1)] with 
$U=U_{Hub}=0$. Varying the spin occupation numbers so that 
$M_{z}\in [0,1]$ we calculate $E_{G}(M_{z})$ whose minimum value 
gives the spin magnetization $M_{z}$ of the system. The stability 
of the self-consistent solutions is very sensitive at larger Coulomb 
interaction strengths, forcing us to keep a relatively small values of 
the interaction parameter $U$. Because of the spin 
rotational symmetry of the Hamiltonian, one limits the numerical 
calculations only to positive spin magnetization 
($n_{\uparrow}\geq n_{\downarrow}$). 

Increasing the amplitude of disorder the spectrum of noninteracting 
Hamiltonian $H_{0}=t\sum_{<i,j>,\sigma}c_{i\sigma}^{\dagger}c_{j\sigma}
+\sum_{i,\sigma}\omega_{i}n_{i,\sigma}$ evolves from weakly to strongly 
localized regime while the eigenvalue statistics exhibits a crossover 
between Wigner and Poisson distributions \cite{zharekeshev}.
As a consequence the distribution function of neighboring level 
spacing undergoes a continuous crossover from Wigner surmise 
$P_{W}(s)=\pi s/2 \cdot \exp(-\pi s^{2}/4)$ with the variance 
$\delta(s)=\sqrt{4/\pi-1} \simeq 0.52$ toward Poisson 
$P_{P}(s)=\exp(-s)$ with $\delta(s)=1$. ($s$ is the dimensionless
level spacing measured in the unit of the mean level spacing
$\Delta =\langle S\rangle$).
This scenario is also 
available when the Hubbard interaction is taken into 
account. For nonzero spin magnetization, the whole spectrum of 
$H^{eff}$ consists of two distinct sequences of the two spectra 
of the $H_{\uparrow}$ and $H_{\downarrow}$. The spectrum of 
$H_{\uparrow}$ depends on the site occupation number 
$n_{\downarrow}(i)$ [c.f. Eq.\,(6)] that will become zero at large 
$U_{H}$ (when $M_{z}=1$ and $n_{\uparrow}=n$),
conserving the eigenstates degree of localization as in the
noninteracting case. Meanwhile, the spectrum of $H_{\downarrow}$ 
depends on the site occupation number $n_{\uparrow}(i)$ (whose  
average over disorder configurations will become
equal to $n/L^{2}$) and moves up at a rate of 
$\langle V_{Hub}^{\downarrow}\rangle_{disorder}=U_{Hub}\cdot n/L^{2}$.
The spectrum of $H_{\downarrow}$ becomes delocalized as a counterpart 
of $V_{Hub}^{\downarrow}(i)$ against the disorder potentials 
$\omega(i)$. In Fig.\,1 we show the differences in the two
sequences of spectrum, $H_{\downarrow}$ states being more delocalized 
compared to $H_{\uparrow}$ states at the same $W$.
The variance  $\delta(s)\simeq 0.7$ in Fig.\,1(a) for $W=10$ as for 
the noninteracting spectrum in the inset of Fig.\,2(c).
In Fig.\,2(c) the variance $\delta(s)\geq 0.8$ for a system size 
$L^{2}=8\cdot 9$ with disorder amplitude $W=20$.
One notes that for a finite system the discreteness of the energy 
spectrum does not allow for a real Poisson distribution
and the level repulsion characteristic to Wigner surmise is present 
even at the critical point of Anderson transition \cite {fyodorov}.
The effect of the Coulomb interaction ($U\not=0,U_{Hub}=0$) over the 
spectrum properties concerns the localization effect of long range 
interaction for $\xi \ge L$ \cite{berkovits}, 
the edge localization of the occupied Hartree levels \cite{vali} 
suggested also in \cite{levit} and increasing gap near
$\epsilon_{F}$ \cite{levit}. In the large lattice size limit the 
Coulomb interaction induces a delocalization process by the meaning
of Poisson-Wigner transition of level spacing distributions \cite{song}.

\begin{figure}
\includegraphics[scale=0.7]{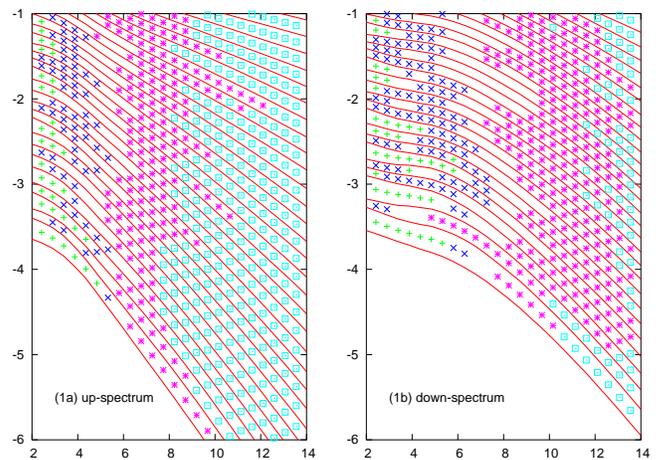}
\caption{The eigenspectrum of $H_\uparrow$ (a) and $H_\downarrow$ (b) 
versus the disorder amplitude $W$ for a system $L^{2}=8\cdot9$, 
$U_{H}=4$, $U=0$ and the electron number $n=8$.
The mark symbols between the eigenvalue lines show different values
of the variance of the neighboring level spacing $\delta (s)$. From left
to right they indicate: $\delta (s)<0.5$ (crosses), 
$\delta (s)\in[0.5,0.54]$ (times), $\delta (s)\in[0.6,0.7]$ (stars) 
and $\delta (s)>0.7$ (boxes). The averages are made over
an ensemble of $500$ configurations.}
\end{figure}
\begin{figure}
\includegraphics[scale=0.7]{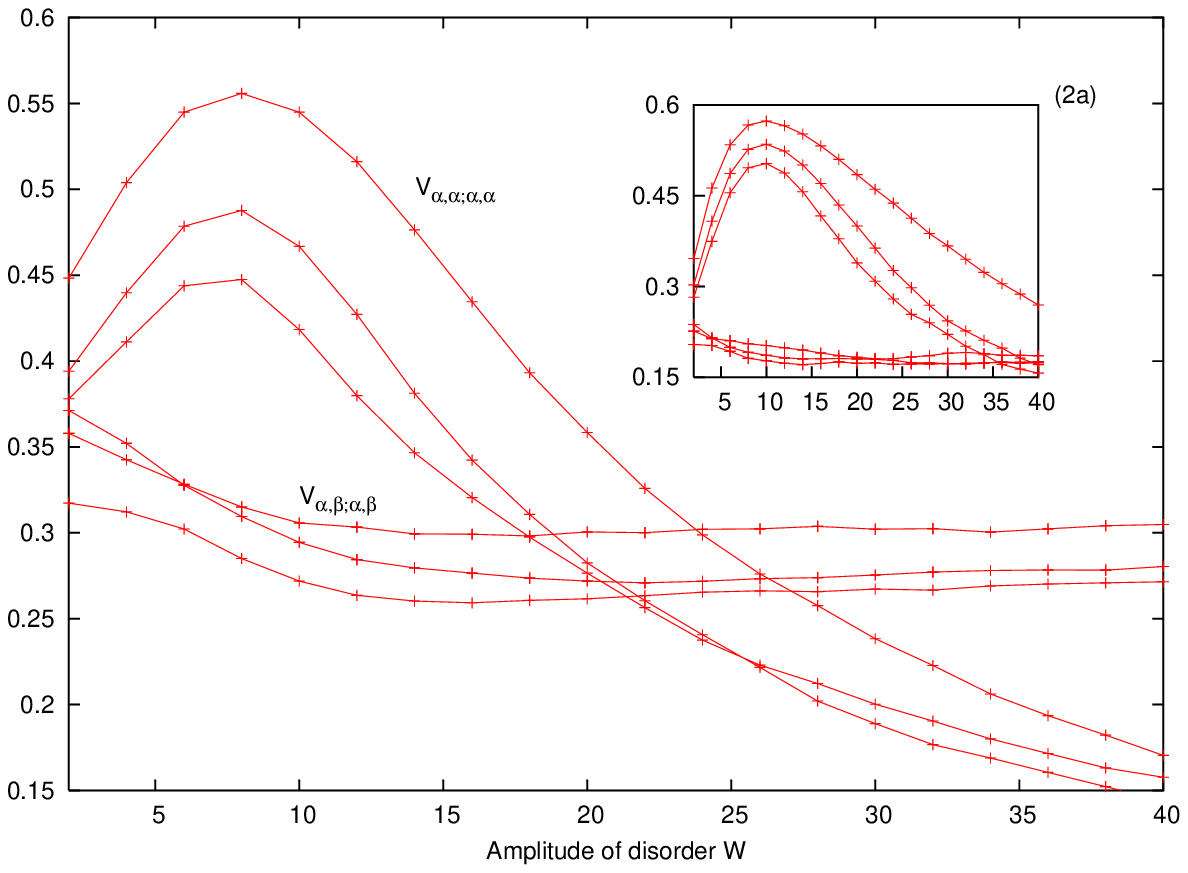}
\includegraphics[scale=0.7]{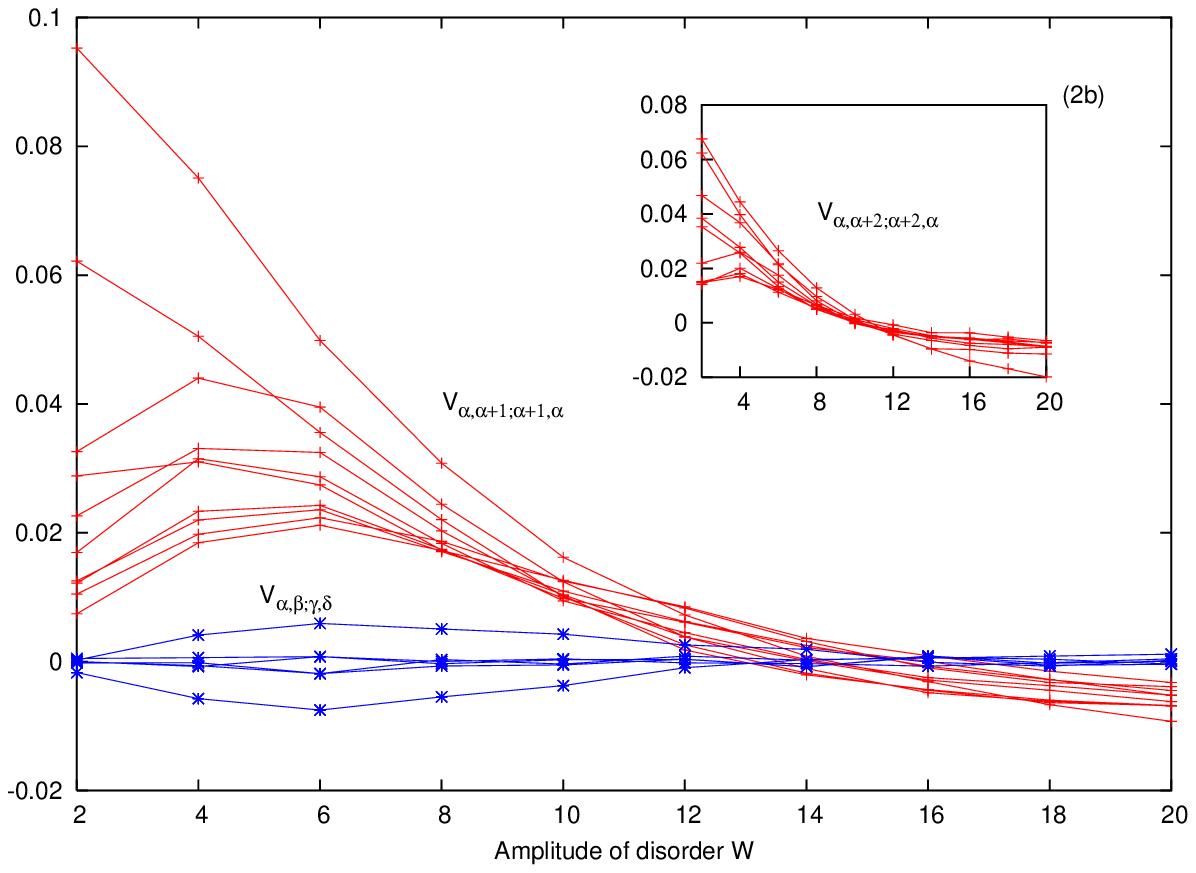}
\includegraphics[scale=0.7]{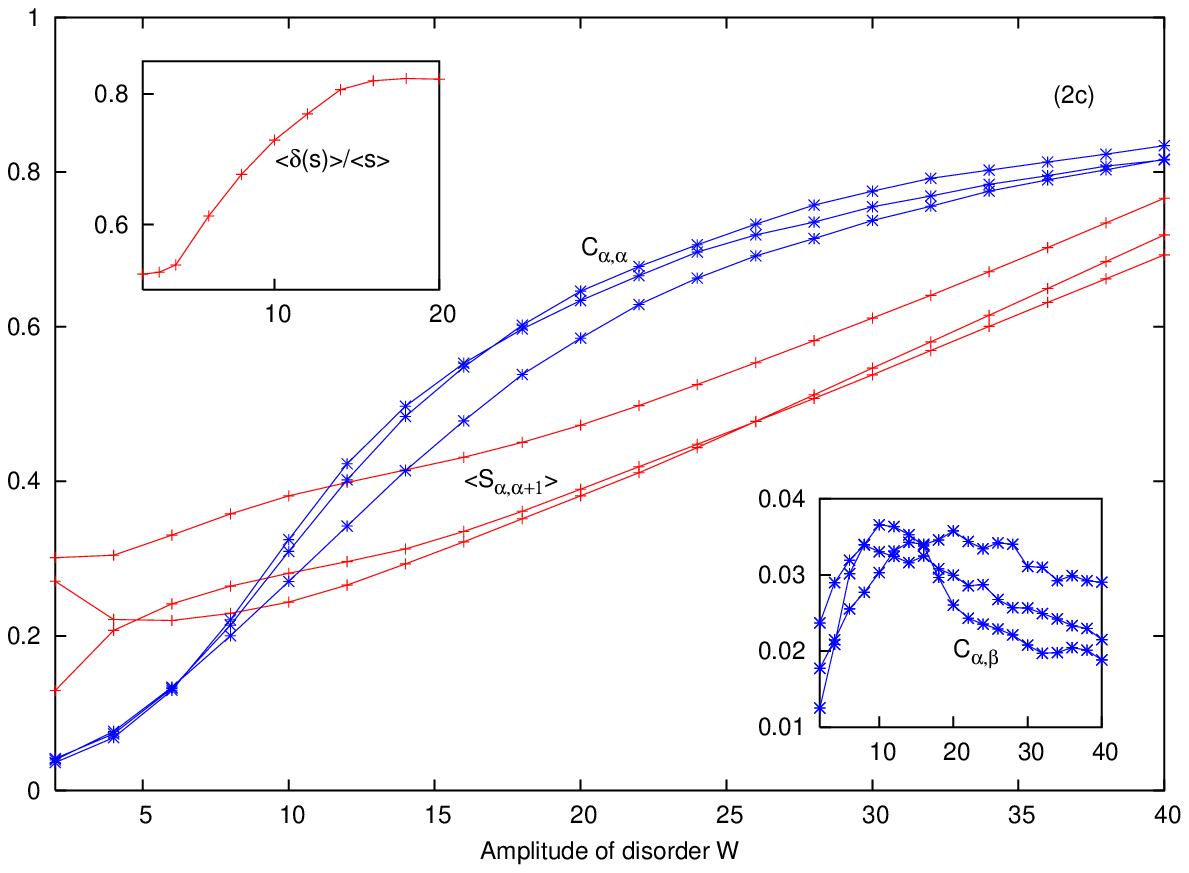}
\caption{The mean values of the interaction matrix elements versus 
amplitude of disorder $W$. $L^{2}=8\cdot 8$.
(a) Direct matrix elements $V_{\alpha\alpha}^{\alpha\alpha}$ with 
$\alpha=1,2,3$ and $V_{\alpha\beta}^{\alpha\beta}$ with $\alpha,
\beta=1,2,3$; (in the inset are represented the same matrix elements 
for $L^{2}=14\cdot 14$).
(b) Exchange matrix elements $V_{\alpha,\alpha+1}^{\alpha+1,\alpha}$ 
with $\alpha=1\ldots 10$ and the off diagonal matrix elements
$V_{\alpha\beta}^{\gamma\delta}$ with $\alpha\beta,\gamma,\delta=1,2,3$.
In the inset $V_{\alpha\alpha+2}^{\alpha+2\alpha}$ for the same system.
(c) Hubbard matrix elements $C_{\alpha\alpha}$ and level spacing
$S_{\alpha\alpha+1}=\epsilon_{\alpha+1}-\epsilon_{\alpha}$. In the 
bottom inset the off diagonal Hubbard elements $C_{\alpha\beta}$ with 
$\alpha\not=\beta$. $\alpha,\beta=1,2,3$. In the top inset the variance
of neighboring level spacing shows the transition from extended regime 
($\delta(s)=0.52$) to localized one ($\delta(s)\to 1$).
The averages are made over 1500 configurations.}
\end{figure}
\begin{figure}
\includegraphics[scale=0.7]{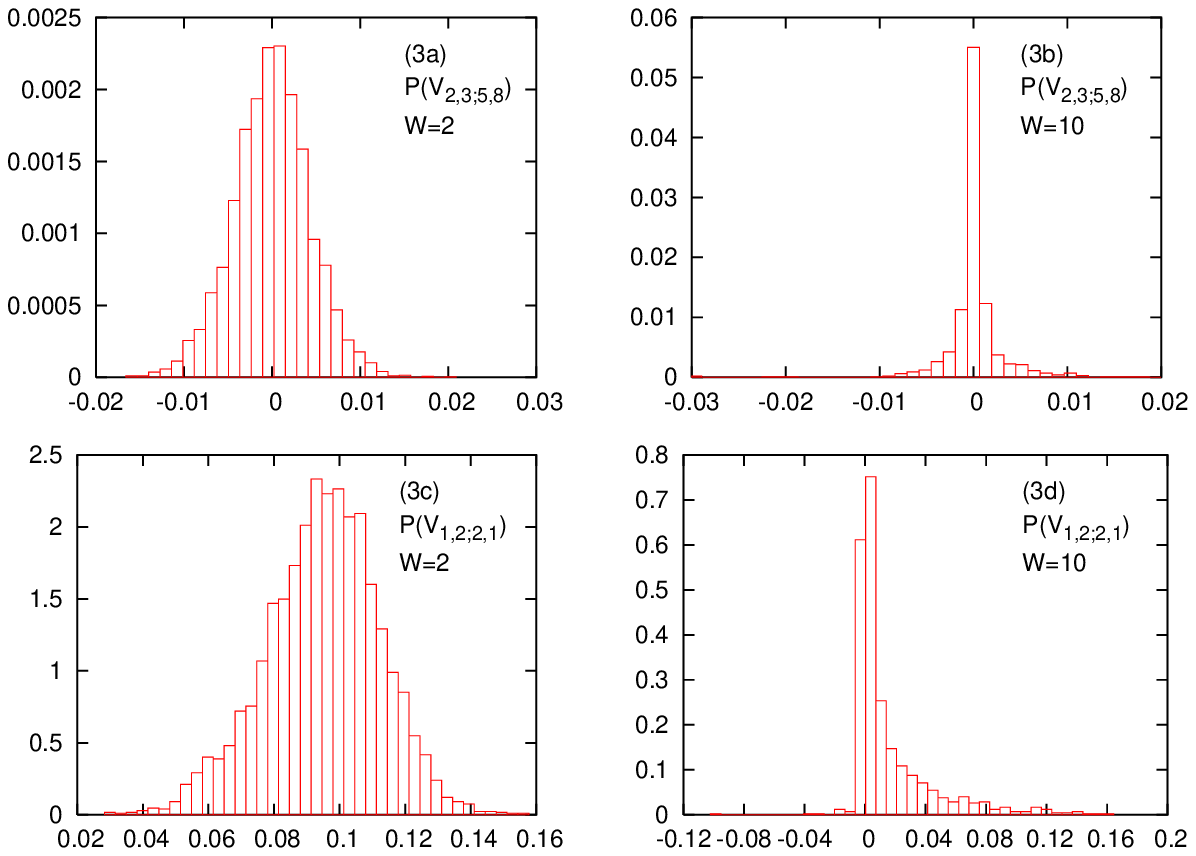}
\includegraphics[scale=0.7]{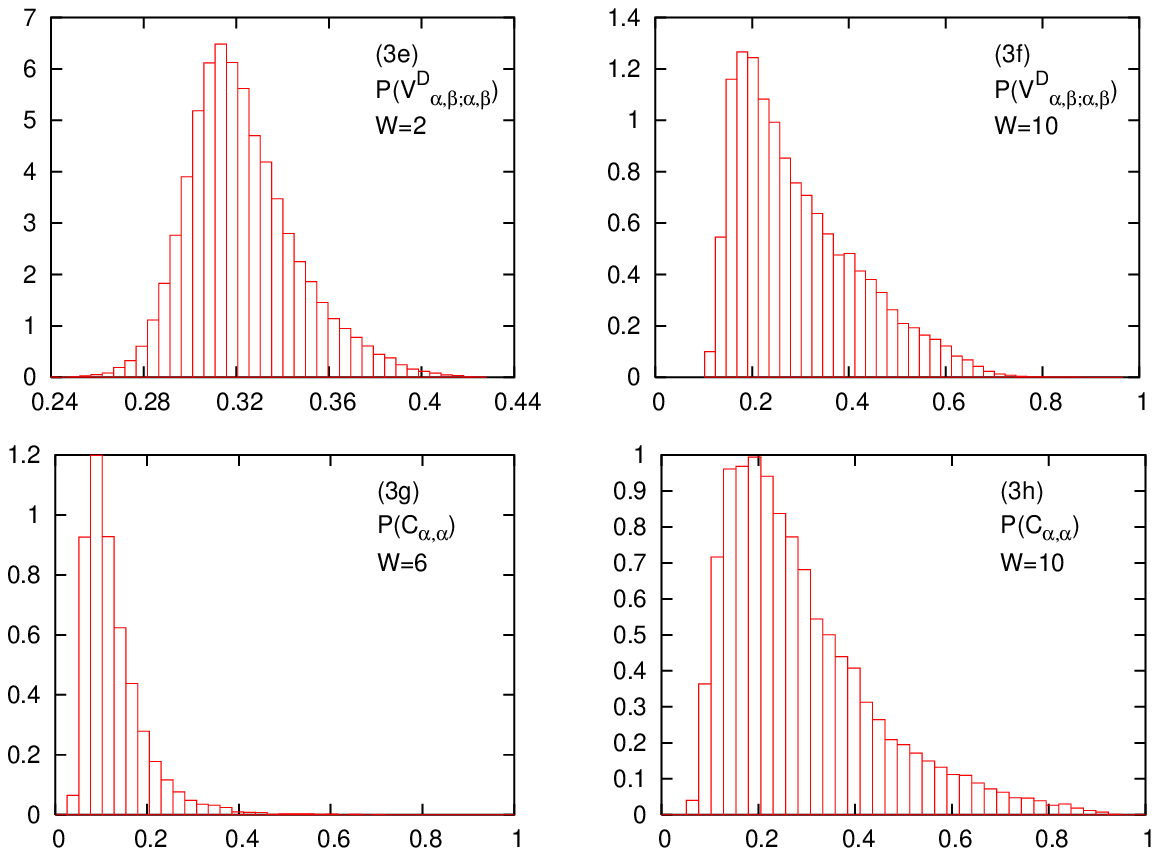}
\caption{The probability density distributions of the interaction matrix
elements. The system dimension $L^{2}=8\cdot 8$. The elements involved 
and the disorder amplitude are marked in every figure.
In (e \ldots h) we represented the distributions of direct and Hubbard 
interaction matrix elements with any $\alpha,\beta \in [1,11]$. For 
$W=10$ all distributions (except for the off-diagonal element) are 
asymmetric with the long tails toward the values larger then the 
disorder averages.
The histograms are represented so that the calculated integral
 be equal to the mean value of the corresponding element.}
\end{figure}

\section {Interaction Matrix Fluctuations}

The Coulomb interaction matrix elements (CME)
$V_{\alpha,\beta}^{\gamma,\delta}$ from Eq.\,(8) and 
the Hubbard matrix elements (HME) $C_{\alpha,\beta}^{\gamma,\delta}$ 
from Eq.\,(9) are calculated for disorder functions in different 
localization regimes. To this purpose, we use the eigenvectors 
$\phi_{\alpha}$ ($\alpha=1\ldots L^{2}$) generated by the 
noninteracting Hamiltonian $H_{0}$ for different degrees of disorder 
characterized by $W$. 
For a 2D lattice with $L^{2}=L_{x}\cdot L_{y}=8\cdot 9$ we have 
identified three 
disordered regimes: (\textit{i}) extended, when $\delta(s)\simeq 0.52 \pm 0.01$ 
for $W\in [1,4]$ (Fig.\,2(c)); (\textit{ii}) strongly localized regime with 
$\xi< L$ and Poisson like repulsion eigenvalues ($\delta(s)> 0.8$ and
$W>15$); (\textit{iii}) weak localization as intermediate for $W\in [4,15]$ 
when the localization length $\xi$ crosses the system length $L$. The 
identification of these three regimes of localization is made for the
purpose of analyzing numerical results presented in this work 
and to help in discussing the features of the interaction 
matrix elements values. When transport properties are involved, 
extended and weakly localized regimes have a common meaning, usually 
addressed as a metallic regime with $\xi> L$ and $g\ge 0.5$ (where
$g$ is dimensionless conductance). In the following we discuss
the main properties of the interaction matrix elements in
various disordered regimes.

%direct direct matrix elements

A. The disorder average of the CME with equal indexes 
$\langle V_{\alpha,\alpha}^{\alpha,\alpha}\rangle$ depends on the 
eigenvectors correlations $\langle |\phi_{\alpha}^{2}(i)|*|
\phi_{\alpha}^{2}(j)|\rangle$ with $i\not=j$. It has large values in 
the extended and weak localization regimes where the correlations of 
the same eigenvectors are proportional to $1/g$ for large 
distances \cite{blanter} and decreasing to zero when the states became 
strongly localized on singular 
lattice points $\phi_{\alpha}(i)=\delta_{i,i_{\alpha}}$ (Fig.\,2(a)).
We note that these matrix elements together with the Hubbard term 
$C_{\alpha,\alpha}^{\alpha,\alpha}$ give the whole interaction energy
of the two electrons (with opposite spin) in the state $\alpha$ 
(c.f. Eq.\,(7)). In contrast, the direct matrix elements 
$\langle V_{\alpha,\beta}^{\alpha,\beta}\rangle$ with $\alpha\not=\beta$
are decreasing when the disorder is increased, keeping constant values 
for strong disorder (Fig.\,2(a)). In this case, assuming that the two states
$\phi_{\alpha}$,$\phi_{\beta}$ are strongly localized in the
two random points $i_{\alpha}$,$i_{\beta}$ we obtain 
$\langle V_{\alpha,\beta}^{\alpha,\beta}\rangle_{config}
=\langle\sum_{i_{\alpha}\ne j_{\beta}}\frac{1}{|i_{\alpha}-
j_{\beta}|}\rangle_{r.p.}$, where the last average is made over all 
possible random pairs of points $(i_{\alpha},i_{\beta})$.
For a 2D system with $L^{2}=8\cdot 8$ and vanishing boundary conditions 
$\langle\sum_{i_{\alpha}\ne j_{\beta}}\frac{1}{|i_{\alpha}-
j_{\beta}|}\rangle_{r.p.}=0.319$ and for $L^{2}=14\cdot 14$ it is equal to 
$0.194$ and these values are close to the averages reached by 
$\langle V_{\alpha,\beta}^{\alpha,\beta}\rangle_{config}$ in Fig.\,2(b).
%exchange matrix elements

B. The exchange matrix elements
$\langle V_{\alpha,\beta}^{\beta,\alpha}\rangle$ and the 
off-diagonal matrix elements $\langle V_{\alpha,\beta}^{\gamma,\delta}
\rangle$ (with $(\alpha,\beta)\not=(\gamma,
\delta)$) in Fig.\,2(b) have maximum values in the weak localization 
regime for $W\simeq 5.5\pm 0.5$ when $\xi > L$.
This is the region where the eigenvector correlations are large.
For uncorrelated states in the strongly localized regime, the 
off-diagonal matrix elements are decreasing to zero faster than the 
exchange matrix elements.
This is also possible for weakly localized regime when the exchange 
elements are calculated between the uncorrelated states whose 
eigenvalues are well separated.
% by at least the Thouless energy of the system
 In the inset of Fig.\,2(b) the exchange matrix elements 
$V_{\alpha,\beta}^{\beta,\alpha}$ are lower when eigenvalue differences
$S_{\alpha\beta}$ are increased. The negative values in Fig.\,2(b) for 
$V_{\alpha,\beta}^{\beta,\alpha}$ when $1\le \xi \ll L$ are a 
property of the finiteness of the system and the orthogonality 
of the eigenvectors $\phi_{\alpha},\phi_{\beta}$.

C. The Hubbard matrix elements are given by the inverse participation
ratio of a state $\phi_{\alpha}$, $C_{\alpha,\alpha}=\sum_{i}
{\phi_{\alpha}^{4}(i)}$ that evolves from $1/L^{2}$ for extended 
states to $1$ for strongly localized ones (Fig.\,2(c)). The 
off-diagonal terms $C_{\alpha,\beta}=\sum_{i}{\phi_{\alpha}^{2}(i)
\phi_{\beta}^{2} (i)}$ decay to zero for strongly localized states and 
have maximum values in the weakly localized regime where the 
correlations of two different eigenfunctions are proportional to 
$1/g$ \cite{blanter}. 
This suggests that the spin polarization transition when $U_{H}\not= 0$
is larger in the strongly disordered regime. As a characteristic of
a finite system, the saturation of the inverse participation ratio 
($C_{\alpha,\alpha}=1$) corroborated with the linear increase 
of the mean level spacing makes impossible a spin polarization 
transition for very large disorder. It does not take place in the 
thermodynamic limit ($L\to\infty$ and lattice parameter $a=const.$) 
when the neighboring level separation $S\to 0$.

The fluctuations of the interaction matrix elements for different 
disorder ranges are presented in Fig.\,3. For small disorder, the 
probability density distributions of the IME are large and they
typically have a Gaussian shape. The exchange (and off-diagonal) matrix 
element fluctuations become sharper when their mean values become
zero and the inverse participation ratio has larger
fluctuations when disorder is increased, with a shape evolution similar
to increasing the Coulomb interaction strength (compare with Fig.\,5 
in Levit and Orgad\cite{levit}). We notice the asymmetric long tail 
fluctuations of direct, exchange and Hubbard matrix elements (Fig.\,3 
for $W=10$), the last two permitting an increased probability of the spin 
polarization transition even when the Stoner criterion does not
allow it.

%direct matrix elements
\begin{figure}
\includegraphics[scale=0.7]{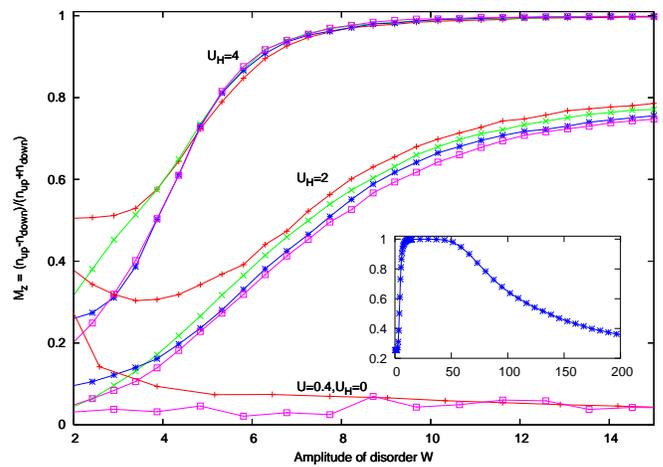}
\caption{The spin magnetization $\langle M_{z}\rangle$ vs. disorder amplitude $W$ for
the filling factor $n/L^{2}=1/9$ and for different interaction strengths
($U_{H}=4,U=0$), ($U_{H}=2,U=0$) and ($U_{H}=0,U=0.4$).
The electron number and the system size are $n=4$ and $L^{2}=6\cdot6$ 
(crosses), $n=6$ and $L^{2}=6\cdot9$ (times), $n=8$
and $L^{2}=8\cdot9$ (stars) and $n=10$ and $L^{2}=9\cdot10$ (boxes). 
In the inset for $U_{H}=4$ and $L^{2}=8\cdot9$ the effect of disorder 
induced spin magnetization disappears when $W>50$. The averages are 
made over $200\sim 1000$ configurations. The large values of $\langle M_{z}\rangle$ 
for  $L^{2}=6\cdot6$ when $W<4$ are due to the degeneracies of the 
spectrum and have no significance for the disorder induced transition.}
\end{figure}
\begin{figure}
\includegraphics[scale=0.7]{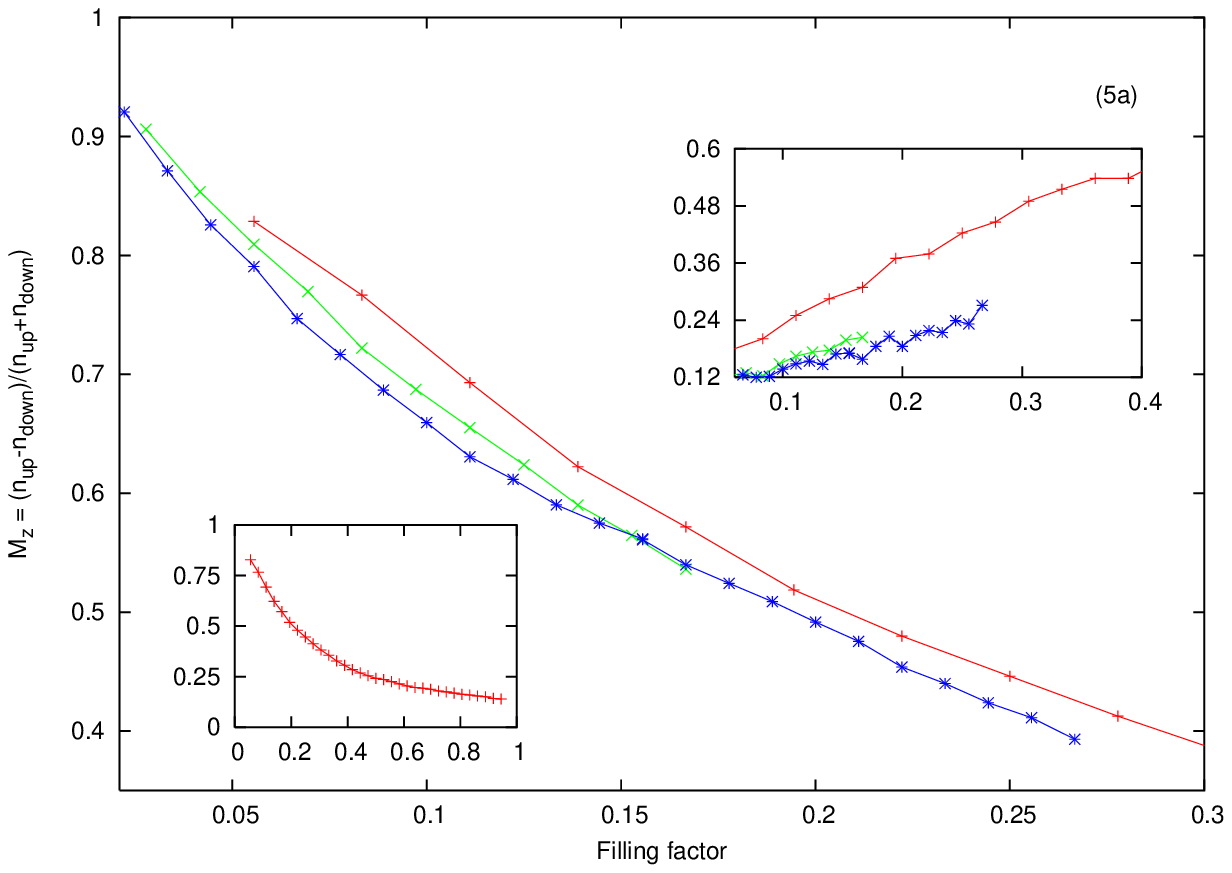}
\includegraphics[scale=0.7]{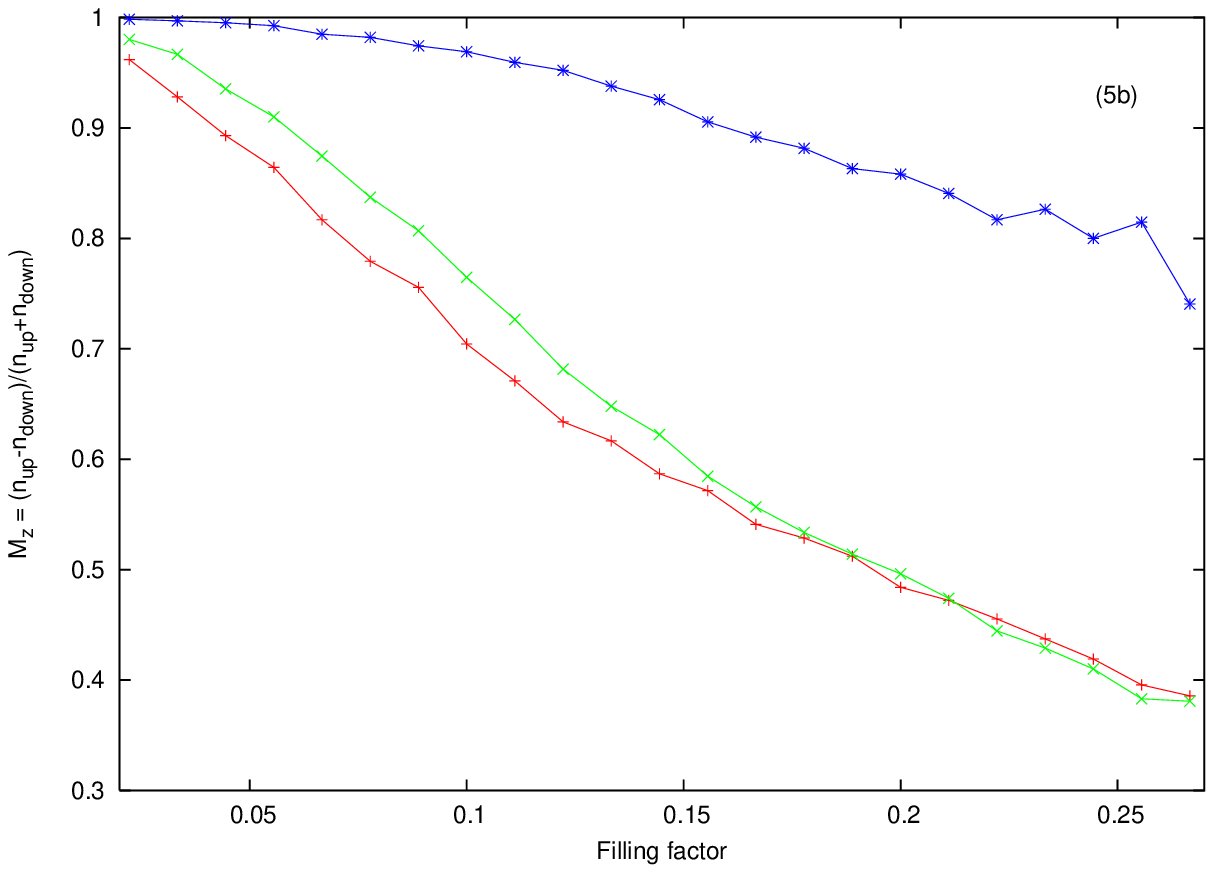}
\caption{The spin magnetization $\langle M_{z}\rangle$ vs. the filling factor 
$n/L^{2}$.
(a) The interaction is $U_{H}=2,U=0$ and disorder $W=10$. In upper 
inset the transmittance of the last occupied level. The three set of 
lines are calculated for $L^{2}=6\cdot6$ (crosses), $L^{2}=9\cdot10$ 
(times) and $L^{2}=9\cdot10$ (stars). In down inset $\langle M_{z}\rangle$ is decreasing 
to zero at $n/L^{2}\propto 1$.
(b) For $W=10$ and $L^{2}=9\cdot10$ the interaction strengths are
$U_{H}=2,U=0.1$ (crosses), $U_{H}=2,U=0.2$ (times) and
$U_{H}=3,U=0.3$ (stars). The averages are made over $200\sim 1000$ 
configurations.}
\end{figure}

\section {Disorder enhanced spin polarization}
The formula of $E_{G}$ given in Eq.\,(7) for the ground state energy
written as an unrestricted Slater determinant [Eq.\,(2)] means that we 
omit the off-diagonal matrix elements from expansion of $E_{G}$.
In Fig.\,2(b) the largest values of the off-diagonal elements in the 
weak localization regime ($V^{off}\simeq 0.008\pm 0.01$) 
are much lower than the mean level spacing ($\delta (S) \simeq 0.2 \pm 0.1$
in Fig.\,2(c)) providing us a good approximation in neglecting 
them when the Coulomb interaction strength $U \ll 5$.
The Stoner criterion for itinerant magnetism is related to the exchange
energy gain by an antisymmetric spin polarized two-body wave function 
that compensates the reduction of the kinetic energy. With our ansatz, in
the regime of extended metallic waves it is expected that the
large exchange matrix elements (especially the elements
$V_{\alpha\alpha+1}^{\alpha+1\alpha}$) can favor a singlet-triplet 
transition when $U\cdot V^{ex}\simeq S$. For a 2D system with 
$L^{2}=8\cdot 9$, $\langle S\rangle\simeq 0.2$ and $\langle V^{ex}\rangle\simeq 0.05$
(in the weak localization regime for $W=6$)  
the interaction strength $U$ must be at least $4$. 
Besides, the enhanced interaction due to the increased 
return probability exhibits a finite temperature partial spin 
polarization \cite {andreev}. By increasing the disorder, the exchange 
matrix elements have positive long tail fluctuations at
values larger than the disorder averages (Fig.\,3(a)) compensating the 
level repulsion of neighboring levels (and favoring also a less 
probable singlet-triplet transition). This will not be possible for 
strongly localized states when any overlap integral is equal to $0$.
%In \cite{milovanovic} the local magnetic moments are because of localized electron and
%interaction between them.
In Fig.\,4 we can see this scenario for a small value of the Coulomb 
interaction $U=0.4$. Difficulties in solving the self-consistent 
equations [Eqs.\,(3-7)] for larger values of the interaction strength 
$U$ prevent us from obtaining reasonable polarization transition
when $U_{H}=0$. The small values of $M_{z}$ in this case represent the 
rare events in the disorder configuration space 
when the fluctuations of $V^{ex}$ are larger than level repulsion.

The disorder induced spin polarization was previously 
obtained \cite{benenti} for the periodic boundary conditions with
$U=U_{H}$ and for a truncated many-body Hilbert space.
This is possible for nonzero Hubbard interaction strength $U_{H}$
when the evolution of $S_{z}$ is increased by disorder
(different curves in Fig.\,4). The full polarization regime
($M_{z}=1$) for a large value of $W$ seems to be stable for the 
constant filling factor $n/L^{2}$ when the system size is increased. 
As it was stressed before, for linearly increasing level spacing at larger 
disorder values, the spin alignment will be turned off when 
$U_{H}\cdot C_{\alpha,\alpha}=U_{H}<S_{\alpha,\alpha+1}\simeq W/L^{2}$
(see inset of Fig.\,4). In Fig.\,5 we calculate the evolution of 
$M_{z}$ varying the 2D electron density ($n/L^{2}$) for $W=10$. When 
the electron number is increased the system shows a smooth decrease
in $M_{z}$. At the same time, the transmittances calculated with
Landauer formula for electrons at Fermi level show the features of 
'insulator-metal' transition (in the inset of Fig.\,5(a) the 
transmittances are enhanced by a factor of about $4$).
As a function of system properties ($W$ and $U_{H}$) the spin 
magnetization can decrease from full or partially spin alignments. 
Our data show a possible transition toward a fully polarized state 
when the electron number $n$ is decreased, suggestive of and
similar to experimental results of Vitkalov {\it et al}. 
\cite{vitkalov} and Shashkin {\it et al}.\cite{shashkin}. This 
takes place only for the values of interaction/disorder parameters 
that allow for a total spin alignment at low electron density 
(see Fig.\,5(b)).

\section {Conclusion}
In this work we have studied the interplay between the disorder
and interaction effects with regard to the spin magnetization of
correlated electrons on a finite 2D lattice.
Our numerical calculations suggest that spin polarization occurs
for a finite 2D system with a model of disorder both for extended and 
localized regimes. The main results we derive are listed below.\\ 
(\textit{i}). The level repulsion of
neighboring states in the extended (and weak localization) regime 
is balanced by the large values of exchange energy or by their long tail
asymmetric fluctuations when the system is gradually localized and the 
eigenfunctions become (extremely) sparse. When the exchange elements 
decrease towards zero (increasing $W$) the spin polarization is not 
allowed unless strong Hubbard repulsion is turned on $U_{H}\not= 0$.\\
(\textit{ii}). The disorder induced spin polarization is calculated for 
a finite 2D strongly correlated system and the data are self scaled in 
the limit of $n/L^{2}=constant$. In the framework of unrestricted HF 
orbitals this is due to the increased separation of up and down 
sequences of the spectrum. When the full polarization is assigned, the 
short range Hubbard interaction will have no localization effect over 
the occupied orbitals while the opposite spin orbitals become 
increasingly delocalized. For a finite system the disorder induced 
spin polarization will be cut off when the linearly increasing level 
spacing exceeds the saturation value of inverse participation ratio.\\
(\textit{iii}). When the number of electrons $n$ decreases the 2D 
system can show a transition to a fully spin-polarized state. 
This, however, takes place only when the 
disorder ($W$) and interaction ($U_{H}$) strengths are properly
tuned to allow for a ferromagnetic spin alignment in the
system at low densities.

\begin{acknowledgments}
This work was supported by CNCSIS and CERES.
M.\,N. acknowledges support from NATO-TUBITAK.
B.\,T. is supported by TUBITAK, NATO-SfP, MSB-KOBRA001, and TUBA.
\end{acknowledgments}

%\begin{thebibliography}{99}

\end{document}